\documentclass[a4paper,10pt,twocolumn]{article}
\usepackage{authblk}

\usepackage{graphicx}
\usepackage{amsmath,amssymb}
\usepackage{hyperref}
\usepackage{lineno}
\usepackage{caption}
\usepackage{siunitx}
\usepackage{float}

\begin{document}

\title{\vspace{-2cm}Heralding efficiency and brightness optimization of a micro-ring resonator via tunable coupling}

\author[1]{Nathan~Moses}
\author[1]{Marcus~J.~Clark}
\author[1]{Alex~S.~Clark}
\author[1]{Siddarth~K.~Joshi}
\author[1,2,*]{Imad~I.~Faruque}

\affil[1]{Quantum Engineering Technology Labs, H. H. Wills Physics Laboratory and School of Electrical, Electronic and Mechanical Engineering, University of Bristol, Tyndall Avenue, Bristol, BS8 1FD, UK}
\affil[2]{School of Mathematical and Physical Sciences, University of Sheffield, Hicks Building, Sheffield, S3 7RH, UK}
\affil[*]{i.faruque@sheffield.ac.uk} 

\date{}

\maketitle

\begin{abstract}
Efficient and bright single photon sources on photonic chips are key to scaling quantum technologies. Spontaneous four wave mixing in micro-ring resonators creates excellent narrowband and tunable photon sources. We experimentally demonstrate the optimization of heralding efficiency and brightness by tuning the coupling of the pump, signal and idler modes into the resonator whilst operating in a pulsed pump regime. We observe a maximum detected pair rate of over 93,000~coincidences per second in a moderately over-coupled regime, alongside a high intrinsic idler heralding efficiency of 97.87$\pm$8.97\% when operating close to maximal over-coupling. The measured dependence on coupling strength is in strong agreement with theoretical predictions, experimentally validating the predicted trade-off between brightness and heralding efficiency. 
\end{abstract}

\section{Introduction}

Quantum technology promises novel computing paradigms, unbreakable secure communication, and enhanced sensing. A leading platform for demonstrating these technologies is photonics, as photons have long coherence times and can encode quantum information in many degrees of freedom~\cite{OBrien2009-fo}. Photonic integrated circuits (PICs) have been used to demonstrate state-of-the-art experiments making use of inherent properties such as tight confinement of light, small form factor and high nonlinearity~\cite{Wang2020, Pelucchi2021-wo}. An important resource of PICs used for quantum applications is the heralded single photon source (HSPS) which utilises nonlinear optical processes such as spontaneous four-wave mixing (SFWM) or spontaneous parametric down conversion (SPDC). The HSPS is pivotal in efficiently generating entangled photon states for photonic quantum technologies~\cite{imog2025}. In SWFM two pump photons from a bright laser source are annihilated to simultaneously create two time-correlated photons, signal and idler, that are detuned from the pump in frequency, with the process obeying energy and momentum conservation. Due to the time correlation, the detection of one of these photons flags the presence of the other photon, creating a heralded single photon source (HSPS). The heralding efficiency is the probability of detecting one photon in a pair given that the other photon has been detected. The detected photon-pair generation rate of a HSPS is referred to as its brightness.  

Various forms of integrated HSPSs have been demonstrated as entanglement sources for quantum networks ~\cite{Appas2021-hm, Zheng2023-rk, Jiang2025} with micro-ring resonators showing promise in terms of scalability ~\cite{Steiner2023-kr, Wen2022-au, Tagliavacche2025}. Micro-ring resonators as ultrabright sources have been implemented using novel platforms, dispersion engineering, source multiplexing and hyper-entanglement \cite{Trevor21, Pang2025, Gianini2026, Wang:25}. Brightness has an impact on both performance and scalability in quantum networking as it defines the maximum possible rate at which entanglement can be shared. In the same regard, the heralding efficiency plays a crucial role in quantum networks as it defines the success rate of any one entangled state being shared. Their combination is often seen as the secret key rate (SKR) in a QKD system \cite{Zo2025, Ecker2021}. Photonic molecules and interferometrically coupled resonators have shown \>90\% heralding efficiency but at the expense of dismantling the regularity of the frequency comb structure of a single conventional resonator ~\cite{Burridge:23}. Another interferometrically coupled design that uses a main cavity and an auxiliary ring shows potential in optimizing heralding efficiency and brightness~\cite{Pagano2024-da} and has been demonstrated to provide a purity of 98.67\% ~\cite{Borghi2024-qa}. Ring resonator HSPSs have also been shown as a preferential source for fusion-based quantum computation ~\cite{Alexander2024-od, Bartolucci2023-xs}. A low heralding efficiency reduces the success of fusion gates, resulting in more attempts required to build the resources used for computation. Efficiency tolerances of fusion gate fidelity have been investigated when regarding emitters \cite{Chan2025}. It is evident that maximising brightness and heralding efficiency of ring resonators can have a strong beneficial impact. 
\par\vspace{-\parskip}
Previous works have related the impact of bus-resonator coupling to ring resonator HSPS performance. One study characterised multiple rings with varying coupling coefficients and observed increased photon-pair generation at certain coupling gaps~\cite{Guo2018-ca}. Other investigations use a resonator structure with a dual asymmetric Mach-Zehnder interferometer (AMZI) coupling to demonstrate increased brightness and heralding efficiency~\cite{Wu2022-vo, Wu2020-fl, Tison2017-xx}. This work differs from these studies by showing stronger conformity with theory by operating in a pulsed pump regime and having uniform reconfigurability of the resonator quality factor (Q-factor) across pump, signal and idler photons which is favourable in applications that involve producing multiple photon-pairs that are wavelength multiplexed across a broadband frequency range, such as quantum networking. 

\begin{figure*}[t]
    \centering
    \includegraphics[width=\linewidth]{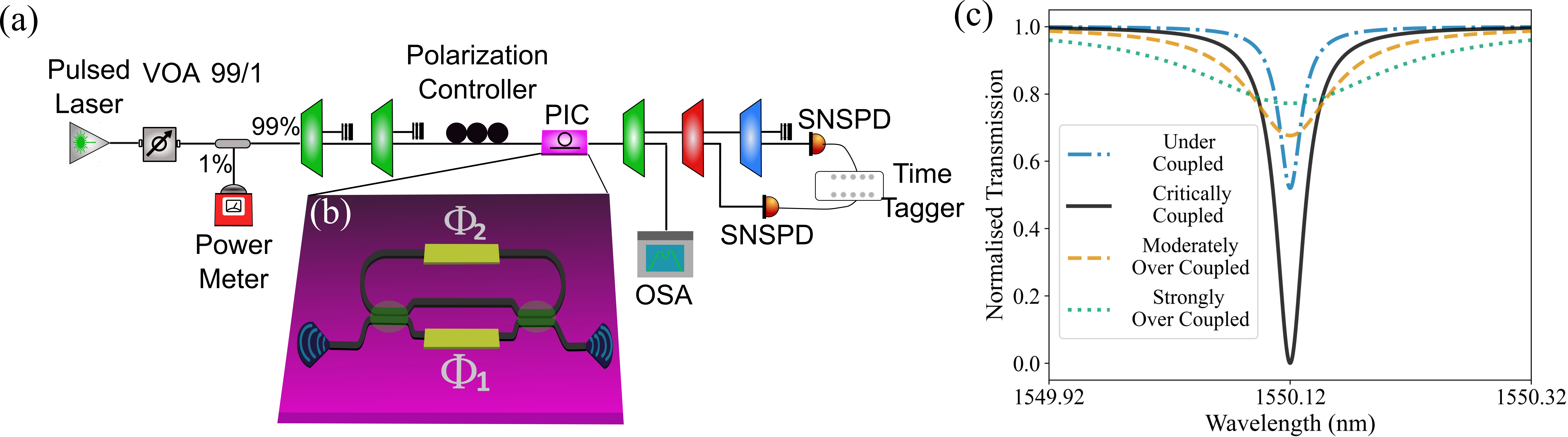}
    \caption{
    Experimental setup for ring resonator characterisation.
    (a) Optical setup used for device characterisation.
    (b) Schematic of the micro-ring resonator with tunable coupling.
    (c) Lorentzian resonance profiles in different coupling regimes.
    }
    \label{fig:Setup}
\end{figure*}

Theoretical investigations suggest that by tuning the coupling of light into a conventional micro-resonator, heralding efficiency can be optimized to a value of 100\% (assuming no losses within the ring) when coupling to the ring is strongest; in reality this value is lower ~\cite{Vernon:16,Sipe2, Bonneau2016}. Theory predicts that the highest brightness is obtained in a moderately over coupled regime and that there is a trade off between brightness and heralding efficiency when going to stronger over coupled regimes.
This work experimentally validates what is predicted in theory ~\cite{Vernon:16,Sipe2, Lukens:26, Sipe2} by demonstrating the optimal coupling regions that maximize brightness and heralding efficiency. 
\section{Experimental Setup}
The experimental setup is shown in Fig.~\ref{fig:Setup}(a). A PriTel femtosecond fiber laser (FFL) with a 50~MHz repetition rate and spectral width of 158~pm is used as the pump. The wavelength is tuned to 1550.12~nm such that it aligns with channel 34 of the ITU C-band. The input pump power is controlled using a variable optical attenuator (VOA).  A 99:1 ratio beam splitter is used so the 1\% tap can monitor the pump laser power. The 99\% output is passed through two cascaded 100~GHz bandwidth channel-34 dense wavelength division multiplexing (DWDM) filters to reduce background noise at the signal and idler wavelengths. As a result, only photons generated from the SFWM process are detected. A fibre polarization controller (FPC) is used to align the input polarization to the grating couplers and waveguides which are optimized for the transverse electric (TE) mode. 

The chip, shown schematically in Fig.~\ref{fig:Setup}(b), contains a ring resonator SFWM photon-pair source. The ring is developed on a 250~nm SOI platform, fabricated by the SiPhotonIC foundry. The silicon layer is 250~nm thick, with a bottom oxide layer thickness of 2~$\mu$m and top oxide thickness of 1~$\mu$m. The resonator and bus waveguides are fully etched strip waveguides chosen to be 450~nm wide such that only single-mode operation of TE polarisation is permitted. The circumference of the ring is chosen to be 699~$\mu$m to match the free spectral range (FSR) of the ring to the 100~GHz spacing of the ITU C-band grid. 

To allow tunability of the coupling from the bus waveguide to and from the ring, we employ a Mach-Zehnder interferometer (MZI)~\cite{Faruque2023-wr}. To change the phase in one MZI arm, a voltage is applied to a thermo-optic phase shifter ($\Phi_1$). Sweeping the voltage changes $\Phi_1$ allowing a range of coupling regimes to the ring to be accessed. Some examples of ring resonances seen in transmission through the bus waveguide are shown in Fig.~\ref{fig:Setup}(c). Raw oscilloscope resonance traces and fitted traces can be found in Supplement Fig S1. and Fig S2. respectively. A full table of extracted resonant parameters as  $\Phi_1$ is tunes can be found in Supplement Table S1. Critical coupling is achieved for a coupling voltage of $\sim1.45$~V, with under/over coupled regimes existing at lower/higher voltages. A second thermo-optic phase shifter ($\Phi_2$) enables tunability of the central ring resonance wavelengths.

After the light is coupled out of the chip a channel 34 DWDM filter is used to reject any residual pump and channel 40 and channel 28 DWDM filters are used to select the signal and idler channels respectively. Single Quantum superconducting nano-wire single photon detectors (SNSPDs) along with a time tagger (Swabian) are used to detect photons and measure correlation between signal and idler detection events.

\section{Results}

To characterise the photon-pair generation and extract the heralding efficiency and brightness for each coupling regime, signal and idler singles counts and coincidence counts were collected for a given MZI voltage over a range of on-chip pump powers. The following equations can be used to describe the SFWM process as a function of pump power~\cite{Bonneau2016}:

\begin{equation}
 C_{s,i}(P) = \eta_{s,i}\bigl(\gamma_{\mathrm{eff}} P^{2} + \beta P\bigr)
 (1-\delta P^{2}) + DC_{s,i},
 \label{eq:1}
\end{equation}\begin{equation}
 CC(P) = \eta_s \eta_i \gamma_{\mathrm{eff}} P^{2}
 (1-\delta P^{2}) + ACC(P).
 \label{eq:3}
\end{equation}

\noindent where $C_{s,i}$ denote the singles counts for the signal and idler photons, and Eq.~\ref{eq:3} describes the coincidence counts ($CC$) as a function of pump power. The $DC$ and $ACC$ parameters refer to the dark counts and accidental counts respectively, and the terms linear in power model the photon counts that are not generated by SFWM in the resonator, such as Raman scattering or leaked pump light. The approximation of the $ACC$ rate can be found in Supplement Eq. S1. Key parameters are the extrinsic heralding efficiencies of the signal and idler photons $\eta_{s,i}$ and the intrinsic generation efficiency $\gamma_{\text{eff}}$, given in units of Mpairs/s/mW$^{2}$, which factors in multiple physical processes such as the $\chi^{(3)}$ of the material, the mode volume of the ring, and the field enhancement of the pump, signal, and idler modes. It also accounts for the acceptance of pump light entering the cavity through the spectral overlap of the pump and ring resonance, as well as the satisfaction of the phase-matching condition. There is a depreciative term with coefficient ${\delta}$ that accounts for the power-dependent degradation of the generation efficiency which encapsulates nonlinear losses such as two-photon absorption (TPA) and associated free carrier absorption (FCA)~\cite{Husko2013-wl, Engin2013-ua}. As these processes are based on the loss of two pump photons, $\delta$ scales with $P^{2}$ resulting in a quadratic depreciation in the generation efficiency.  Fitting was also performed at low pump powers to negate the effects of nonlinear losses and can be found in Supplement Fig. S4. Using the model that includes nonlinear losses, the experimental data can be fit to Eqs.~(\ref{eq:1}) and (\ref{eq:3}). Examples of fitted singles and coincidences counts can be found in Supplement Fig. S3. A full table of extracted parameters at each setting of $\Phi_{1}$ can be found in Supplement Table S2.

We consider the intrinsic heralding efficiency of the signal and idler photons as the escape efficiency. This is the probability that the photons couple to the bus waveguide after they have been created within the cavity. To extract the escape efficiency, all losses from the ring until detection are accounted for as 

\begin{equation}
    \eta_{k_\text{intrinsic}} = \eta_{k_\text{escape}} =\frac{\eta_{k}}{\eta_{\text{gc}}\eta_{\text{channel,k}}\eta_{\text{det}}}\,,
    \label{eq:7}
\end{equation}

\noindent where $k = \{s,i\}$, $\eta_{\text{gc}}$ is the grating coupler efficiency (58.2\%) and $\eta_{\text{det}}$ represents the detector efficiency (88\%), with both assumed to apply equally to the signal and idler photons. The channel efficiencies, $\eta_{\text{channel,s}}=35.7\%$ and $\eta_{\text{channel,i}}=35.4\%$ include all losses from the chip to the detector system. 
For each bus-ring coupling regime where photon-pair statistics have been measured and efficiencies extracted, the predicted outcomes for brightness and heralding efficiency are found given the measured Q-factor of the resonances. To compare to theory, we use the following parameters~\cite{Vernon:16}

\begin{equation}
    \Gamma = \frac{\omega}{Q_{ext}}\,, \hspace{0.8cm}
    M = \frac{\omega}{Q_{int}}\,, \hspace{0.8cm}
    \Gamma + M = \frac{\omega}{Q_{loaded}}\,,
    \label{eq:8}
\end{equation}

\noindent where $\omega$ is the angular frequency of the mode. $\Gamma$ is used to model the coupling rate between the bus waveguide and ring and is defined by the resonance extrinsic quality factor, $Q_{ext}$. $M$ models the scattering loss that arises from intrinsic losses of the ring itself, represented by the intrinsic Q-factor, $Q_{int}$, that is independent of any coupling to a bus waveguide. $Q_{loaded}$ is the experimentally measured Q-factor and comprises both $\Gamma$ and $M$ . It is the relationship between $\Gamma$ and $M$ that determines the coupling strength between the bus waveguide and resonator. Ref.~\cite{Vernon:16} suggests that the maximum coincidence rate occurs when $\Gamma = 4M$, while other works have proposed that the maximum is achieved at $\Gamma = 2M$ \cite{Lukens:26, Sipe2}. One study further demonstrated that the expected ratio varies on the pump bandwidth \cite{Wu2022-vo}. Our experimental results are consistent with the modelled expression presented in \cite{Vernon:16}, which is therefore used as the theoretical model for brightness. These studies also predict that heralding efficiency is maximised when coupling is maximised, which implies $\Gamma \gg M$. These expressions can also be found in Supplement Eq. S2. Using the measured intrinsic and extrinsic quality factors, the theoretical model can be compared with our experimental measurements.

\begin{figure}[!t]
    \centering
    \includegraphics[width=\columnwidth]{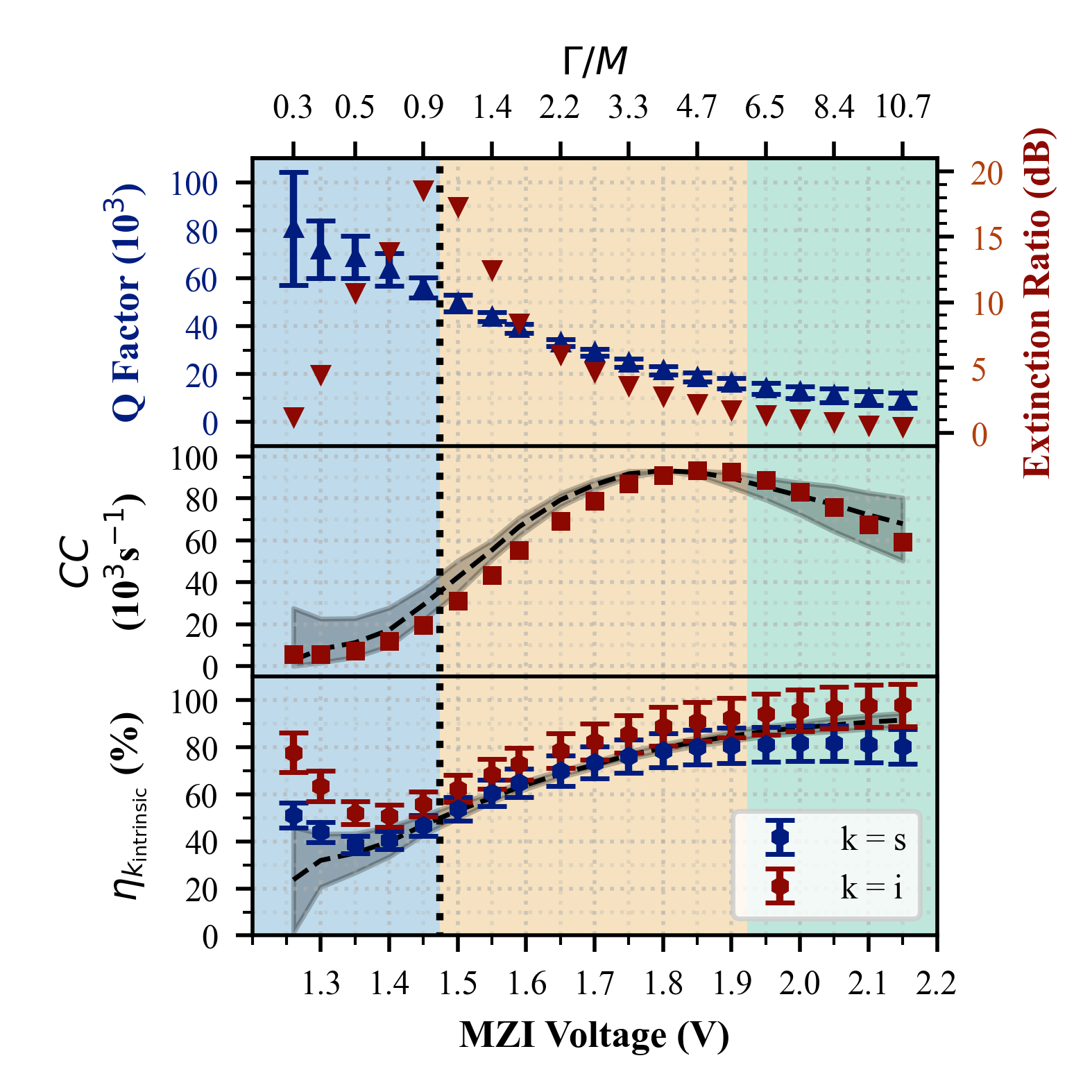}
    \caption{
     The optical response to ring resonator coupling variation. The top plot shows the Q-factor (blue triangles) and extinction (red inverted triangles) of the ring resonances. Q-factor error comes from fitting of resonances. The middle plot shows the detected coincidences per second (red squares). The lower plot shows the show extracted intrinsic heralding efficiencies $\eta_s$ (blue circles) and $\eta_i$ (red circles). Error bars arise from uncertainty in system losses and goodness of fit. Black dashed lines show theory predictions with shared region as the bounds. Shaded regions indicate the under-coupled regime (blue), the moderately over-coupled regime (yellow), and the strongly over-coupled regime (green). The black dotted line is critical coupling.
    }
    \label{fig:Data}
\end{figure}

The measured Q-factor and extinction of the ring resonances for varying coupling are shown in Fig.~\ref{fig:Data}. The Q-factor increases with decreased coupling, while the extinction peaks at critical coupling. The value of $M$ takes into account the resonator cavity length, round trip transmission and group index of the waveguide. We find a round trip transmission of 94.4\% from the highest extinction resonance. This round trip transmission comprises the waveguide loss, quoted by the foundry to be 2.5~dB/cm, with the remaining losses assumed to come from the MZI directional couplers, $\sim0.05$~dB/coupler. The measured counts and parameters extracted from our fitting can now be compared to theory, shown in Fig.~\ref{fig:Data}. The measured coincidence rate shows good agreement with theory. The peak at a $\Phi_{1}$ voltage of 1.85~V corresponds to a coupling regime of $\Gamma = 4.7M$, which slightly differs from the expected maximum point of $\Gamma = 4M$. The observed discrepancy may stem from the broader resonances present at stronger coupling, which exhibit increased tolerance to pump–resonance misalignment and group velocity dispersion (GVD), consequently mitigating DWDM related losses and increasing the detected pair rate. With pump alignment, this is regarding the relative alignment of the micro-ring resonances with respect to the pump wavelength as the $\Phi_{1}$ and $\Phi_{2}$ phases are tuned. Small shifts in resonance detuning modify the intracavity pump power spectral density, which in turn affect the degree to which the SFWM energy-matching condition is satisfied. Relative misalignment between coupling regimes would provide some non-conformity with the predicted trend. In the over-coupled regime the heralding efficiencies agree well with theory, increasing as $\Gamma \gg M$. It is evident that $\eta_{i} > \eta_{s}$. At each coupling strength it is expected that the signal and idler photons share the same escape efficiency such that $\eta_{i} = \eta_{s}$. This constant discrepancy could be attributed to a relative loss that is not accounted for, such as unequal fibre coupling losses or differing detector efficiencies. The earlier plateau of the $\eta_{s}$ efficiencies may be due to biased ring coupling of the idler mode compared to signal at these strongly over-coupled regimes due to improper centering of the pump resonance. There could also be a wavelength-dependent phase difference from the signal and idler modes due to chromatic dispersion in the directional couplers. This could cause the signal mode to maximally over couple at a lower voltage setting of the $\Phi_{1}$ phase shifter. The largest discrepancy that arises from both methods lies in the under-coupled regimes. Coupling points at MZI voltages below 1.4~V show an increase in heralding efficiency as the coupling of the ring to the bus waveguide gets weaker. One explanation for this is that at such weak coupling regimes the detected photon generation rate (PGR) from the cavity is similar to the PGR arising from SFWM occurring solely in the bus waveguide. Thus, the model fits the measured photons counts to two independent SFWM processes that inflate the extracted efficiencies.
\par\vspace{-\parskip}
By comparing the points of maximum brightness and heralding efficiency, it is evident there is a trade-off between heralding efficiency and brightness, with both points existing at different strengths in the over-coupled regime, as predicted by theory ~\cite{Vernon:16, Sipe2, Lukens:26}. This trade-off can be attributed to the balance between the probabilities of photons completing round trips inside the ring and photons escaping the ring into the bus waveguide. Higher Q-factors (weaker coupling regimes) signify longer photon lifetimes in the resonator. For pump photons this increases the likelihood of pair-production. As signal and idler photons also inherit these long resonator lifetimes, the chance of both escaping the resonator diminishes as the Q-factor increases. This is because the photons are more likely to be lost to the scattering modes of the ring than couple back into the bus waveguide. For lower Q-factors (stronger coupling regimes) the pump photons have shorter resonator lifetimes but signal and idler photons have greater chance of escaping the cavity. This provides an increased probability that both signal and idler photons will be detected at the expense of lower pair-production rates. In the moderately over-coupled regime, an optimal balance is achieved between pump coupling into the cavity, intracavity field enhancement, and pump photon lifetime. This balance maximizes the PGR while maintaining sufficiently high escape efficiencies for simultaneous detection of the signal and idler modes. As a result, the detected coincidence rate is higher than in other coupling regimes.

\section{Conclusion}

We experimentally investigated the optimization of heralding efficiency and brightness of photon-pair generation in a single micro-ring resonator by tuning the coupling to and from the bus waveguide to the ring. We used an MZI structure for selecting the coupling coefficients. Our experimental results follow the trends and values predicted by the theory. Maximum brightness is observed at a slightly over-coupled regime and not at the critical coupling regime, and the observed heralding efficiency continues to increase as we continue to over-couple the resonator up to a maximum that is close to unity. This work illustrates the usefulness of such a tunable source design in quantum technology applications such as quantum networking and photonic quantum computing and provides a framework for optimizing micro-ring resonators as heralded single photon sources. 

{\small
\section*{Acknowledgments}

This work was supported by the UK Engineering and Physical Sciences Research Council (EPSRC) grants EP/T001011/1 -- the Quantum Communications Hub and  EP/Z533208/1 -- the Integrated Quantum Networks (IQN) Hub which are both part of the UK national quantum technologies programme. It was also supported by the Integrated Multi-Channel Bell State Generator (Ref. No R2305011) Partnership Resource Fund and the EPSRC new investigator award EP/X039439/1 -- Towards The Quantum Internet: Interconnecting Quantum Networks. A.S.C. also acknowledges support from The Royal Society (URF/R/221019, RF/ERE/210098, RF/ERE/221060). The authors thank John Rarity and Milica Banic for useful discussions.

}



\end{document}